\documentclass{aa}  

\usepackage{graphicx}
\usepackage{txfonts}
\begin{document}

   \title{A generalizable method for estimating meteor shower false positives}


   \author{P.M. Shober
          \inst{1}
          \and
          J. Vaubaillon\inst{1}\fnmsep
          }

   \institute{Institut Mécanique Céleste et de Calcul des Éphemerides, Observatoire de Paris, PSL, 75014, Paris, France\\
              \email{patrick.shober@obspm.fr}}

   \date{Received December 19, 2023; accepted February 29, 2024}

 
  \abstract
   {The determination of meteor shower or parent body associations is inherently a statistical problem. Traditional methods, primarily the similarity discriminants, have limitations, particularly in handling the increasing volume and complexity of meteoroid orbit data.}
   {We aim to introduce a new, more statistically robust and generalizable method for estimating false positive detections in meteor shower identification, leveraging Kernel Density Estimation (KDE).}
   {Utilizing a dataset of 824 fireballs observed by the European Fireball Network, we apply a multivariate Gaussian kernel within KDE and z-score data normalization. Our method analyzes the parameter space of meteoroid orbits and geocentric impact characteristics, focusing on four different similarity discriminants: $D_{SH}$, $D'$, $D_{H}$, and $D_{N}$.}
   {The KDE methodology consistently converges towards a true established shower-associated fireball rate within the EFN dataset of 18-25\% for all criteria. This indicates that the approach provides a more statistically robust estimate of the shower-associated component.}
   {Our findings highlight the potential of KDE, combined with appropriate data normalization, in enhancing the accuracy and reliability of meteor shower analysis. This method addresses the existing challenges posed by traditional similarity discriminants and offers a versatile solution adaptable to varying datasets and parameters.}


   \maketitle
%



\section{Introduction}
The identification and analysis of meteor showers, distinct from the sporadic meteoroid background, is a pivotal aspect of meteor science. The International Astronomical Union (IAU) maintains a comprehensive list of these meteor showers and their associated parent bodies. This list is continuously updated with new entries derived from ongoing meteoroid orbit surveys, emphasizing the dynamic nature of this field \citep{jenniskens2009report,kornovs2014confirmation,rudawska2015independent,jenniskens2016established}. The process of discerning these meteor showers from the background noise hinges on the application of a D-criterion. Initially formulated by \citet{southworth1963statistics}, the $D_{SH}$ is calculated from two sets of orbital elements. This parameter, along with its derivatives, increases with decreasing similarity and is sometimes referred to as a ‘dissimilarity parameter’. \citet{southworth1963statistics} designed their parameter for ease of computation and acknowledged the possibility of valid alternative formulations. One known flaw of this parameter though is the physical unit inconsistency \citep{drummond1980meteor,drummond1981test}.

Variations such as \citet{drummond1981test} D-parameter have been developed. The orbital similarity parameter of \citet{drummond1981test}, akin to the version of \citet{southworth1963statistics}, balances the weights of its four terms and uses angular distances rather than chords. A comparison by \citet{jopek1993remarks} highlighted the overdependence of the \citet{southworth1963statistics} parameter on perihelion distance and \citet{drummond1981test} parameter on eccentricity, leading to the proposal of a hybrid parameter, $D_{H}$. These criteria all quantify the similarity between meteoroid orbits, with lower D-values indicating greater resemblance. Traditionally, a threshold value of \( D_{SH} < 0.2 \) was employed, but lower values have since been used in recent studies \citep{jenniskens2009report,rudawska2015independent, kornovs2014confirmation, jenniskens2016established}. In addition, \citet{valsecchi1999meteoroid} introduced a new distance function based on four geocentric quantities directly linked to observations. This methodology differs from the conventional orbital similarity criteria, like the \citet{southworth1963statistics} criterion, by focusing on variables that are near-invariant with respect to the principal secular perturbation affecting meteoroid orbits. The introduction of this new approach offers a more observationally direct and more accurate means for classifying meteoroid streams. 

Despite its widespread use, the validity of the D-criterion as an absolute measure has been questioned, particularly its reliance on sample size and the potential for misidentifying random associations as genuine meteor showers \citep{pauls2005decoherence,koten2014search,egal2017challenge,vida2018modelling}. Given the rapid expansion of publicly available meteor orbit data, primarily from video meteor networks, there arises a need for more robust methods to assess the statistical significance of the association between meteor showers and their proposed parent bodies. This need is further underscored by the increasing realization that meteoroid orbital elements are often more uncertain than previously thought \citep{egal2017challenge,vida2018modelling}.  Notably, \citet{moorhead2016performance} and \citet{sugar2017meteor} have contributed significantly to the detection of meteor showers by employing the density-based spatial clustering algorithm (DBSCAN) to large datasets of meteor trajectories and orbits observed by the NASA All-Sky Fireball Network and the Southern Ontario Meteor Network. These studies have underscored the necessity for more robust methods in assessing meteor shower associations, especially considering the uncertainties and varying characteristics of meteoroid orbits. 

In response to these challenges, this study introduces a new method to estimate D-criterion false positives, centered around Kernel Density Estimation (KDE), a non-parametric way to estimate the probability density function of a random variable. KDE is advantageous over traditional histograms due to its continuity and flexibility in dealing with various data distributions. Our approach addresses the limitations inherent in traditional methods that rely heavily on the similarity discriminant, a parameter that has been historically used but often leads to ambiguous results. In our approach here, we employ a multivariate Gaussian kernel to model the distribution of sporadic meteoroid orbits. This kernel is defined by a covariance matrix that simplifies the complex relationships between orbital elements into a manageable form. We then generate synthetic samples from this KDE to estimate the sporadic background, subsequently enabling us to assess the likelihood of false positive detections in meteor shower identification. 

This methodology allows for a statistically sound examination of potential false associations between observations and meteor showers or parent bodies, aiming to address the shortcomings of previous methods and provide a more reliable and generalized framework for meteor shower analysis.

\section{Method}

The current state of meteor shower analysis, primarily governed by the similarity discriminant, presents several limitations. The discriminants, while useful, often lead to ambiguous results, especially when the underlying sample size is not adequately considered. This ambiguity is evident in the extensive list of potential parent bodies on the working list of meteor showers\footnote{\url{https://www.ta3.sk/IAUC22DB/MDC2022/Etc/streamworkingdata2022.txt}}, many of which may be spurious associations. Furthermore, the increasing volume of meteoroid orbit data, accompanied by significant uncertainties in their orbital elements, necessitates a more comprehensive method to distinguish between genuine meteor showers and random groupings within the sporadic background \citep{egal2017challenge,vida2017generating}.

To address these challenges, our study employs KDE as a foundational tool for analyzing the orbital distributions of meteoroids. KDE's non-parametric nature allows for flexible and unbiased estimation of density functions, making it particularly suited for handling the diverse and often complex distributions encountered in meteor shower analysis \citep{seaman1996evaluation}. By synthesising large samples from the KDE and comparing these against the observed meteoroid orbits, we can effectively gauge the probability of random coincidences, thereby refining our understanding of meteor shower associations.

Our methodology can be summarized as follows: 
\begin{enumerate}
    \item \textit{Data Collection and Preparation:} Gather a dataset of meteor observations, concentrating on the orbital and geocentric parameters necessary to calculate the $D_{SH}$, $D'$, $D_{H}$, and $D_{N}$ similarity discriminants. Use z-score normalization on data, to prepare for KDE. 
    \item \textit{Kernel Density Estimation (KDE):} Apply KDE to the normalized data to estimate the probability density function of the sporadic meteor background. This involves selecting an appropriate kernel (e.g., Gaussian) and bandwidth, which determines the level of smoothing. 
    \item \textit{Calculation of Dissimilarity Criteria:} Compute various discriminant values (e.g., \(D_{SH}\), \(D'\), \(D_{H}\), \(D_{N}\)) for the dataset. These criteria assess the similarity between observed meteoroid orbits and known meteor showers.
    \item \textit{False Positive Estimation:} Randomly draw $N_{dataset}$ synthetic samples from the PDF for the sporadic population, estimated by the KDE, and un-normalize the parameters for discriminant value calculation. Determine the rate of false positives by calculating the discriminant values for the synthetic sporadic samples, and finding how many have low D-values by chance. 
    \item \textit{True Positive Identification:} Subtract the estimated false positives from the total associations with the observed dataset to determine the true positive total, providing a clearer picture of genuine meteor shower associations.
\end{enumerate}

\subsection{Data}
In this paper we are using the dataset of 824 fireballs observed by the European Fireball Network (EFN) \citep{borovivcka2022data}. The EFN, established in 1963, represents a pioneering effort in the long-term monitoring of fireballs using a network of wide-angle and all-sky cameras. Initially set up in Czechoslovakia and Germany, it has expanded significantly in terms of geographic coverage and technological advancement. The network, primarily based in Central Europe, has undergone several modernizations over the decades, including the transition from mirror to fish-eye cameras and, more recently, the adoption of digital autonomous observatories. These advancements have significantly enhanced the network's capabilities in detecting and analyzing fireballs. The network's data have contributed to the recovery of several meteorites and have provided valuable insights into the physical and orbital properties of meteoroids. 

\subsection{Data Processing}
In this study, we employed a rigorous data normalization and KDE approach to analyze and estimate the sporadic background of meteor observations, no matter the variable of interest. In order to well-utilize a KDE for this purpose, the data needs to be normalized in order to avoid over- or under-smoothing. It ensures that all features in a dataset contribute equally to the analysis. Without normalization, features with larger scales can disproportionately influence the KDE, leading to biased results. For example, if you want to fit a KDE to orbital data, the range of the angular elements will either be 0-180$^{\circ}$ or 0-360$^{\circ}$, whereas the eccentricity will only vary between 0.0-1.0. Without some normalization, the semi-major axis and eccentricity features in the dataset would likely be over-smoothed. 

In this study, we follow a standard practice of normalization by using a Z-score normalization \citep{glantz2001primer}. Z-score normalization, also known as standard score normalization, is a statistical method used to standardize the features of a dataset. It is defined by the formula:
\begin{align*}
    z = \frac{(X - \mu)}{\sigma}
    \label{eq:z_score}
\end{align*}
where $X$ is the original data value, $\mu$ is the mean of the data, and $\sigma$ is the standard deviation. In this process, each feature value is transformed by subtracting the mean of the feature and then dividing by its standard deviation. This normalization process facilitates the conversion of each feature to a scale with a mean of 0 and a standard deviation of 1, thereby rectifying the issue of disparate scales that could lead to over- or under-smoothing in the KDE process. Such standardization is indispensable in our analysis, given the diverse range and nature of the orbital and geocentric parameters under consideration. This transformation is applied individually to each variable and does not alter the relative positioning of individual data points within each feature. Importantly, because Z-score normalization is applied feature-wise, it does not inherently disrupt the correlation structure between features. Furthermore, the reversible nature of Z-score normalization permits the re-scaling of the KDE output to the original data scale, enabling the interpretation and application of the results within the authentic context of the observed meteoroid orbits. Furthermore, it is imperative to acknowledge that while Z-score normalization aids in harmonizing the scale across parameters, it does not alter the underlying distributional characteristics of the data, such as skewness or kurtosis. This initial phase of preprocessing the data was implemented using the StandardScaler method from the scikit-learn library \citep{pedregosa2011scikit}. 

Within this study, we use a normalization process to allow the use of a single bandwidth parameter to generate uniform smoothing across dimensions. However, the methodology proposed by \citet{vida2017generating}, which introduces the use of a diagonal bandwidth matrix (Equation 12 in \citealp{vida2017generating}), represents another alternative that merits consideration. This other approach allows for the individual adjustment of bandwidths for each parameter, tailored to their specific distributional properties, thereby offering a more manual alternative to method implemented here. Thus, if someone wants to smooth certain features more than others, this approach should be utilized instead. However, for demonstrating how smoothing generally effects the shower false positive estimate, our methodology that allows for the use of one bandwidth value is ideal. 

\subsection{Kernel Density Estimation}
Post-normalization, we implemented the KDE, a non-parametric way to estimate the probability density function of a random variable, using the KernelDensity class from scikit-learn. KDE is particularly beneficial in elucidating the underlying structure of the data, especially when the form of the distribution is unknown \citep{silverman2018density}. A KDE is a non-parametric method used to estimate the probability density function of a random variable based on a data sample. It works by placing a kernel, typically a Gaussian function, on each data point and then summing these kernels to create a smooth estimate of the underlying probability density function. This method is particularly useful for approximating unknown distributions and accommodating the multi-modality often present in sparse data \citep{silverman2018density}. Additionally, KDE is known for its flexibility in accurately estimating densities of various shapes, provided that the level of smoothing is appropriately selected \citep{seaman1996evaluation}.

Within this study, we utilize a Gaussian kernel in order to provide enough smoothing to estimate the sporadic meteor component from the dataset. However, before fitting the KDE to the data, all established showers were identified and removed in order to ensure that the false positive rate was not over-estimated. One could use a KDE with a sufficiently large bandwidth to effectively smooth-away all of the shower-related features, however, if the shower component is significant this will reduce the accuracy of the sporadic distribution. If the shower component is in fact dominating the dataset under question, as seen in many datasets (Table 1; \citealp{jopek1997stream}), this method will over-estimate the false-positive rate. In this study, we are using a dataset of 824 EFN fireballs, of which up to 45\% were estimated to be shower-associated \citep{borovivcka2022data}. However, \citet{borovivcka2022data} stated that shower membership listed is obvious for well-defined major showers but many of the minor shower might not even be real. We chose to remove the established major shower components, removing all fireballs with a $D_{N}<0.1$ with an established shower\footnote{\url{https://www.ta3.sk/IAUC22DB/MDC2022/Roje/roje_lista.php?corobic_roje=1&sort_roje=0}}. If the shower-component of a dataset is very large, the well-established showers need to be removed, however, one must do this carefully as an over-removal and under-smoothing of the KDE will conversely result in an underestimate of the meteor shower false positive rate. 

The KDE for a univariate dataset is defined as:
\[
f(x) = \frac{1}{nh} \sum_{i=1}^{n} K\left(\frac{x - x_i}{h}\right)
\]

where:
\begin{itemize}
    \item \( f(x) \) is the estimated density at point \( x \).
    \item \( n \) is the number of data points.
    \item \( x_i \) are the data points.
    \item \( h \) is the bandwidth, a smoothing parameter.
    \item \( K \) is the kernel, a non-negative function that integrates to one and has mean zero.
\end{itemize}

The choice of kernel function \( K \) and the bandwidth \( h \) are crucial.

For multivariate data, the KDE becomes:

\[
f(\mathbf{x}) = \frac{1}{n \, \text{det}(H)} \sum_{i=1}^{n} K\left(H^{-1} (\mathbf{x} - \mathbf{x}_i)\right)
\]

where:
\begin{itemize}
    \item \( \mathbf{x} \) and \( \mathbf{x}_i \) are now vectors.
    \item \( H \) is the bandwidth matrix, generalizing the smoothing parameter to multiple dimensions.
    \item \( \text{det}(H) \) is the determinant of \( H \), normalizing the kernel.
\end{itemize}

The Gaussian kernel \( K_{Gaussian} \), used in this study, is defined as: 

\[
K_{Gaussian}(u) = \frac{1}{\sqrt{2\pi}} e^{-\frac{1}{2} u^2}
\]

where:
\begin{itemize}
    \item \( K(u) \) is the Gaussian kernel.
    \item \( u \) is the standardized variable, calculated as \( \frac{x - x_i}{h} \), where \( x \) is the evaluation point, \( x_i \) is a data point, and \( h \) is the bandwidth.
    \item \( e \) is the base of the natural logarithm.
    \item The kernel integrates to 1 over its domain, conforming to the properties of a probability density function.
\end{itemize}

Or in the multivariate case: 

\[
K_{Gaussian}(\mathbf{u}) = \frac{1}{(2\pi)^{\frac{d}{2}} \text{det}(\Sigma)^{\frac{1}{2}}} \exp\left(-\frac{1}{2} \mathbf{u}^\top \Sigma^{-1} \mathbf{u}\right)
\]

where:
\begin{itemize}
    \item \( K(\mathbf{u}) \) represents the multivariate Gaussian kernel.
    \item \( \mathbf{u} \) is the standardized variable vector, calculated as \( H^{-1} (\mathbf{x} - \mathbf{x}_i) \), where \( \mathbf{x} \) is the evaluation point vector, \( \mathbf{x}_i \) is a data point vector, and \( H \) is the bandwidth matrix.
    \item \( d \) is the number of dimensions.
    \item \( \Sigma \) is the covariance matrix, often related to the bandwidth matrix \( H \).
    \item \( \text{det}(\Sigma) \) is the determinant of the covariance matrix.
    \item \( (2\pi)^{\frac{d}{2}} \text{det}(\Sigma)^{\frac{1}{2}} \) normalizes the kernel to ensure it integrates to 1.
    \item \( \exp \) is the exponential function.
    \item \( \mathbf{u}^\top \) is the transpose of \( \mathbf{u} \).
    \item \( \Sigma^{-1} \) is the inverse of \( \Sigma \).
\end{itemize}

The equations for the KDE and corresponding Gaussian kernel were taken from \citet{hastie2009elements}; please refer to this text for a more detailed description. 

Despite its advantages, there are considerations to be made when using KDE. For instance, the level of smoothing, determined by the bandwidth parameter, must be carefully chosen to avoid under-fitting or over-fitting the data. Here, we consider multiple bandwidths to demonstrate how this parameter affects the estimate of the sporadic meteor population and subsequently the meteor shower false positive rate for each similarity discriminant. 

The utilization of KDE on cyclic or periodic data also warrants careful consideration, especially given the intrinsic challenges posed by such data types. Cyclic parameters, such as the angular elements in meteoroid orbital data, exhibit continuity at their boundaries -- a property that conventional KDE approaches, including those predicated on linear kernels, may not adequately accommodate. This discontinuity at the boundary can lead to misleading density estimates, particularly near the edges of the cyclic range. The process of Z-score normalization, while facilitating the standardization of data scales, does not inherently resolve the cyclic nature of these parameters. Consequently, the application of KDE to unmodified cyclic data, even post-normalization, necessitates a methodological adjustment to ensure that the periodic continuity is preserved and accurately represented in the density estimation.

To address this challenge, specific strategies may be employed, such as the adaptation of KDE with cyclic kernels or the transformation of cyclic data into a format that inherently respects its periodic boundaries. These adaptations are essential for capturing the true density landscape of cyclic parameters, ensuring that the estimations reflect the natural continuity and cyclic behavior inherent to such data. This consideration underscores the importance of selecting appropriate KDE configurations and transformations that align with the data's characteristics, thereby enhancing the accuracy and relevance of the density estimates derived from our analysis. The nuanced handling of cyclic data within KDE highlights the broader theme of methodological adaptability, emphasizing the need for tailored approaches that are sensitive to the unique properties of the dataset under investigation.

\subsection{Dissimilarity Criteria}

A similarity discriminant is a statistical measure used in astronomy to evaluate the similarity between the orbital elements of meteoroids, asteroids, or comets. It has been refined into several versions, each with unique characteristics and calculations. We detail four prominent versions of the D value: $D_{SH}$ \citep{southworth1963statistics}, $D'$ \citep{drummond1981test}, $D_{H}$ \citep{jopek1993remarks}, and $D_{N}$ \citep{valsecchi1999meteoroid}. 

\subsubsection{Southworth-Hawkins Discriminant}
The study by \citet{southworth1963statistics} was the first to introduce a method to identify meteoroid streams. Their approach was to calculate an orbital discriminant ($D_{SH}$) based on the calculated pre-impact orbital elements of the meteors detected by Baker Super-Schmidt meteor cameras. 

The $D_{SH}$ orbital similarity discriminant is defined as: 
\begin{align*}
D_{SH}^{2} = & \left( q_B - q_A \right)^2 + (e_B - e_A)^2 + \left( 2 \sin\left(\frac{I}{2}\right) \right)^2 \\
&+ \left( (e_B + e_A) \cdot \sin\left(\frac{\pi}{2}\right) \right)^2 \\
\text{where} \\
I &= \arccos\left( \cos(i_A) \cdot \cos(i_B) + \sin(i_A) \cdot \sin(i_B) \cdot \cos(\Omega_A - \Omega_B) \right) \\
\pi &= \omega_A - \omega_B \\
&+ 2 \arcsin \left( \frac{\cos\left(\frac{i_A + i_B}{2}\right) \cdot \sin\left(\frac{\Omega_A - \Omega_B}{2}\right)}{\cos\left(\frac{I}{2}\right)} \right) \\
\text{arcsin} &= \begin{cases} 
negative & \text{if } |\Omega_A - \Omega_B| > 180^\circ \\
positive & \text{otherwise}
\end{cases}
\end{align*}


\subsubsection{Drummond Discriminant}

The Drummond discriminant, proposed by \citet{drummond1981test}, is also an orbital discriminant that can be used to differentiate between small bodies and meteors based on their orbital elements. It is expressed as:

\begin{align*}
D'^{2} &= \left( \frac{e_B - e_A}{e_B + e_A} \right)^2 + \left( \frac{q_B - q_A}{q_B + q_A} \right)^2 + \left( \frac{I}{180^\circ} \right)^2 \\
&+ \left( \frac{(e_B + e_A)}{2} \frac{\theta}{180^\circ}\right)^2  \\
\text{where} \\
\Theta &= \arccos\left[ \sin(\beta'_B) \sin(\beta'_A) + \cos(\beta'_B) \cos(\beta'_A) \cos(h'_B - h'_A) \right], \\
\text{with} \\
\beta' &= \arcsin\left( \sin(i) \sin(\omega) \right), \\
h' &= \Omega + \arctan\left( \cos(i) \tan(\omega) \right) + (\cos(\omega) < 0) \cdot 180^\circ.
\end{align*}

This criterion focuses more on the differences in eccentricity \( e \) and perihelion distance \( q \) between two orbits, compared to \citet{southworth1963statistics}.

\subsubsection{Jopek Discriminant}

The discriminant, introduced by \citet{jopek1993remarks}, is a more complex variant that combines elements of \citet{southworth1963statistics} and \citet{drummond1981test} and can be written as:

\begin{align*}
D_{H}^{2} &= \left(e_B - e_A\right)^2 + \left( \frac{q_B - q_A}{q_B + q_A} \right)^2 + \left( 2 \sin\left(\frac{I}{2}\right) \right)^2 \\ 
&+ \left( (e_B + e_A) ( 2 \sin\left(\frac{\pi}{2}\right) \right)^2
\end{align*}

\subsubsection{Valsecchi Discriminant}

The Valsecchi discriminant, $D_{N}$, developed by \citet{valsecchi1999meteoroid}, takes a completely different approach using four geocentric quantities directly linked to meteor observations. This diverges from the traditional discriminant values based on the osculating orbital elements at impact. The proposed approach defines the distance function in a space with dimensions equal to the number of independently measured physical quantities. 

The new variables introduced are:
\begin{itemize}
    \item The modulus of the unperturbed geocentric velocity, \( U \).
    \item Two angles, \( \theta \) and \( \phi \), defining the direction of \( U \) based on \"Opik's theory. These angles are used to define the direction opposite to that from which the meteoroid is observed, considering Earth's gravity effect.
    \item The solar longitude of the meteoroid ($\lambda$) hitting the Earth 
\end{itemize}

\citet{valsecchi1999meteoroid} recommended using \( \cos v \) instead of \( v \), as it is directly proportional to \( -\frac{1}{a} \) (the orbital energy of the meteoroid), making it suitable for the new distance function.

The similarity criterion, \( D_{N} \), is defined as:
\begin{align*}
    D^{2}_{N} &= (U_2 - U_1)^2 + w_1 (\cos \theta_2 - \cos \theta_1)^2 + \Delta\xi^2 \\
    \text{where} \\ 
    \Delta\xi^2 &= \min \left( w_2 \Delta\phi_I^2 + w_3 \Delta \lambda_I^2, \, w_2 \Delta\phi_{II}^2 + w_3 \Delta \lambda_{II}^2 \right) \\
    \Delta\phi_I &= 2 \sin\left(\frac{\phi_2 - \phi_1}{2}\right) \\
    \Delta\phi_{II} &= 2 \sin\left(\frac{180^\circ - \phi_2 - \phi_1}{2}\right) \\
    \Delta \lambda_I &= 2 \sin\left(\frac{\lambda_2 - \lambda_1}{2}\right) \\
    \Delta \lambda_{II} &= 2 \sin\left(\frac{180^\circ - \lambda_2 - \lambda_1}{2}\right)
\end{align*}
and $w_1$ , $w_2$ , $w_3$ are suitably defined weighting factors (all are set to 1.0 here); note that $\Delta\xi$ is small if $\phi_1 - \phi_2$ and $\lambda_1 - \lambda_2$ are either both small or both close to $180^{\circ}$.

\subsection{Estimating the Number of False Positives}

To estimate the number of false positives within a given dataset for a given similarity discriminant, we start by using the KDE to estimate a probability density function (PDF) of the sporadic population. From this sporadic PDF, samples are randomly drawn in batches of $N_{dataset}$, where $N_{dataset}$ is the total number of datapoints in the dataset, i.e., the total number of fireballs in the EFN dataset. 

We employ scikit-learn's \texttt{KernelDensity} class to perform random sampling from a fitted Gaussian Kernel Density Estimation \citep{pedregosa2011scikit}. The sampling algorithm initiates by randomly selecting base points from the dataset used in the KDE fitting, ensuring an equitable chance of selection across data points, unless sample weights are provided, in which case selection probabilities are adjusted accordingly. Upon selecting base points, the algorithm introduces Gaussian noise to each, effectively drawing a random sample from a Gaussian distribution with a mean equal to the base point's coordinates and a standard deviation determined by the KDE's bandwidth parameter. Mathematically, for a base point denoted as $x_i$, a sampled point $x'_i$ is produced according to $x'_i = x_i + N(0, h^2)$, where $N(0, h^2)$ signifies a Gaussian noise component with mean 0 and variance $h^2$, and $h$ represents the bandwidth. This procedure, repeated for each base point, generates samples that reflect the density estimated by the KDE, with the bandwidth parameter serving as a critical smoothing factor, influencing the dispersion of generated samples around the base points and thereby controlling the smoothness of the estimated distribution. Ultimately, this method yields a set of points distributed according to the original dataset's estimated probability density function, as modeled by the Gaussian KDE, facilitating the exploration of continuous data distributions and leveraging the Gaussian kernel's inherent properties for applications such as Monte Carlo simulations and synthetic data creation. For more information regarding the scikit-learn's \textit{KernelDensity} class, please refer to the source code on GitHub\footnote{\url{https://github.com/scikit-learn/scikit-learn/blob/9e38cd00d/sklearn/neighbors/_kde.py#L35}}.

This process is done 100 times to get good statistics on the likelihood of a chance association for a dataset of that size, as it has been shown previously that the number of false positives changes with sample size \citep{southworth1963statistics}. 

For each random sample of size $N_{dataset}$ drawn from the sporadic meteor PDF, the similarity discriminant is calculated. In this study, we calculate the $D_{SH}$, $D'$, $D_{H}$, and $D_{N}$ values for every random draw and every meteor shower in the IAU Meteor Data Center's Established Shower list\footnote{\url{https://www.ta3.sk/IAUC22DB/MDC2022/Roje/roje_lista.php?corobic_roje=1&sort_roje=0}}. Assuming the estimated PDF accurately reflects the distribution of sporadic sources within the dataset, we interpret the count of samples whose similarity discriminant values fall below a specified threshold (referred to as the D-criterion) as our estimate of false positives. This means that, within the context of our analysis, any sample from the sporadic source distribution that yields a similarity discriminant value lower than this threshold is considered a false positive. This threshold-based approach allows us to quantify the likelihood of mistakenly identifying sporadic sources as members of a significant pattern or group when, in reality, they are not, based on the statistical properties of the dataset under investigation. Additionally, by taking a Monte Carlo approach, uncertainty can be placed on the false positive estimate.

\section{Results}

Applying a KDE to fireball network observations in order to estimate the meteor shower false positives and, inherently, the true positives, shows a lot of promise. Here, we have applied the KDE to the observations to estimate the sporadic distribution for the variables involved in the D discriminant calculations for $D_{SH}$, $D'$, $D_{H}$, and $D_{N}$. The parameters change between these D-values, but the KDE false positive method seems to provide a generalizable way of estimating the shower component despite this change. 

\subsection{KDE Smoothing}
\begin{figure}
	\includegraphics[width=0.9\columnwidth]{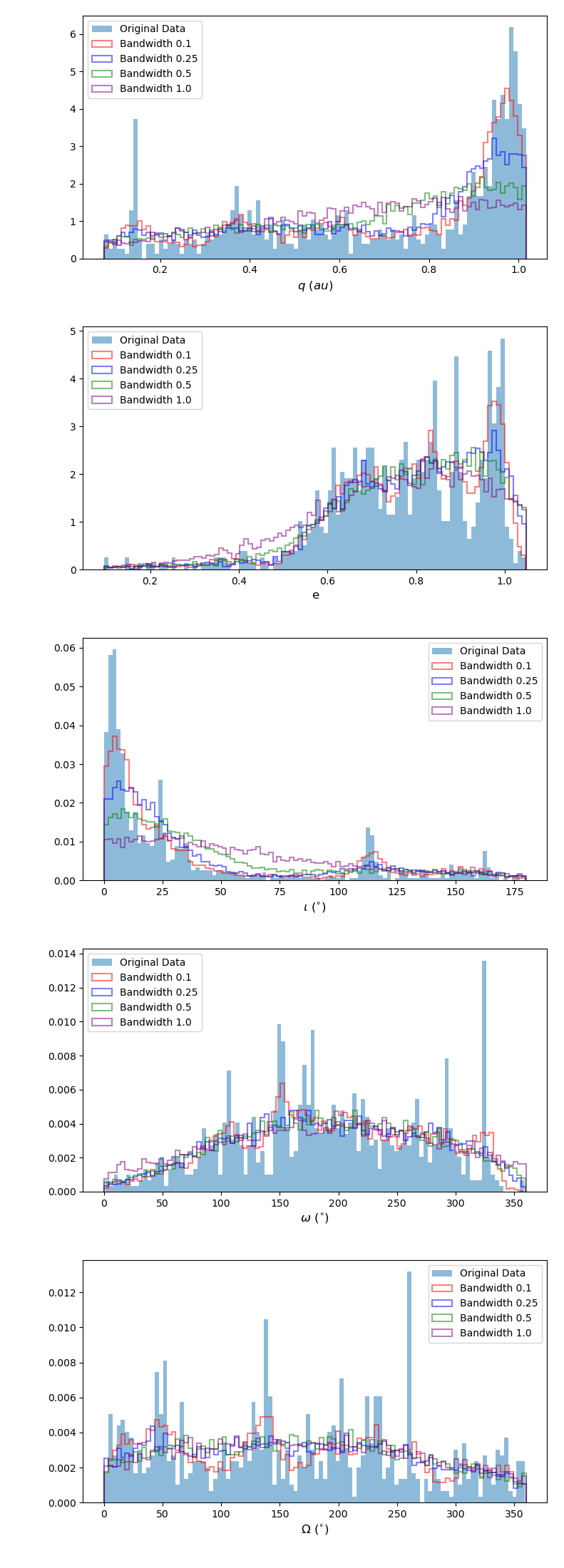}
    \caption{Histogram of observed distribution of EFN orbital parameters used to calculate $D_{SH}$, $D'$, and $D_{H}$ with curves denoting the PDF estimated by a KDE with various bandwidth values.}
    \label{fig:orbit_bandwidths}
\end{figure}

\begin{figure}
	\includegraphics[width=0.9\columnwidth]{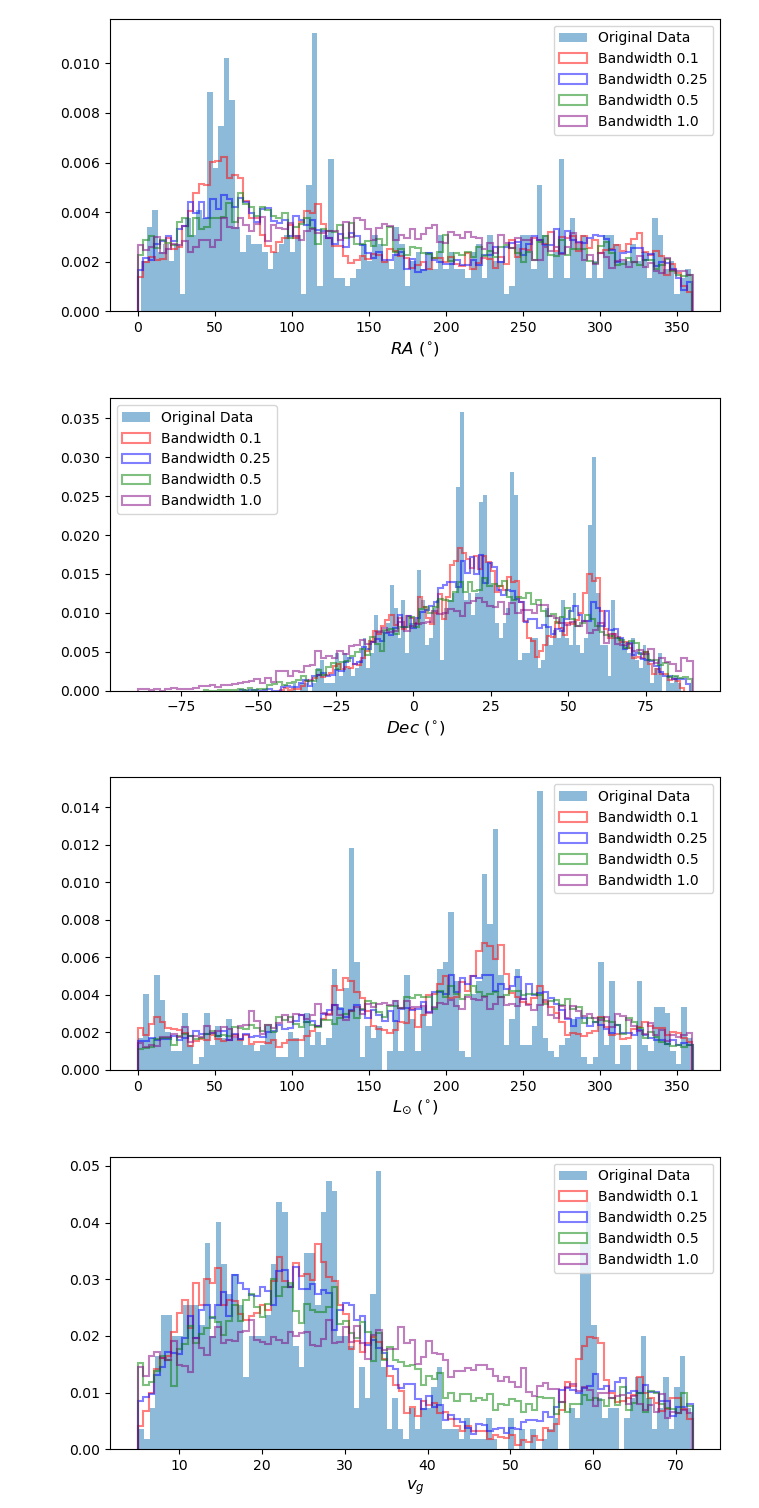}
    \caption{Histograms of observed distribution of EFN geocentric parameters used to calculate $D_{N}$ with curves denoting the PDF estimated by a KDE with various bandwidth values.}
    \label{fig:geo_bandwidths}
\end{figure}

Due to our z-score normalization of the data, we can apply the same level of smoothing to different parameters whose range and magnitudes vary significantly using one bandwidth parameter value. As seen in Fig.~\ref{fig:orbit_bandwidths} and Fig.~\ref{fig:geo_bandwidths}, we have applied multiple KDEs with increasing bandwidth values to the observed parameters that are used to calculate the discriminants. Fig.~\ref{fig:orbit_bandwidths} compares the PDFs of the orbital parameters used to calculate $D_{SH}$, $D'$, and $D_{H}$ to the observed distributions. Whereas, Fig.~\ref{fig:geo_bandwidths} does the same for the geocentric parameter used to calculate $D_{N}$. We applied a KDE with four different bandwidth values [0.1, 0.25, 0.5, 1.0] in order to demonstrate the effect on the false positive estimate. As the bandwidth values increase towards 1.0, the features of the original measured distribution become less pronounced. 

Although we tested values ranging up until 1.0, this bandwidth is too large and starts to remove some of the likely sporadic features in the dataset as well. Using cross-validation and grid search techniques to ascertain the optimal bandwidth for KDE with a Gaussian kernel gives a reasonable value of 0.22 for the EFN dataset. This `optimum' bandwidth is found through cross-validation, specifically a 5-fold variant. Cross-validation involves partitioning the data into five subsets, or 'folds'. This partitioning is key to ensuring that the evaluation of the bandwidth's effectiveness is comprehensive and unbiased. Each subset is alternately used as a testing set while the remaining are amalgamated to form a training set. In each iteration, one of the five folds is designated as the 'test set', and the remaining four folds are combined to form the 'training set'. The 'training set' in this context refers to the subset of data on which the KDE is applied to estimate the density function. The 'test set' is then used to evaluate the effectiveness of this density estimation. This process is iterated five times, ensuring each subset serves as the testing set once. A grid-search approach iterates over a predefined range of bandwidths, in this instance from 0.1 to 1.0, segmented into 30 intervals. The performance of the KDE model for each bandwidth value is evaluated, and the optimal bandwidth is defined as the one striking a balance between over-smoothing, which obscures pertinent details of the data distribution, and under-smoothing, which introduces excessive noise.

This method to determine the `optimal' bandwidth can work if the shower components reassemble noisiness in the data. However, if the showers make up a much larger component, the major established showers will need to be removed during a pre-processing step to adequately estimate the underlying sporadic distribution. 

\subsection{False Positive Estimate}
\begin{figure}
	\includegraphics[width=\columnwidth]{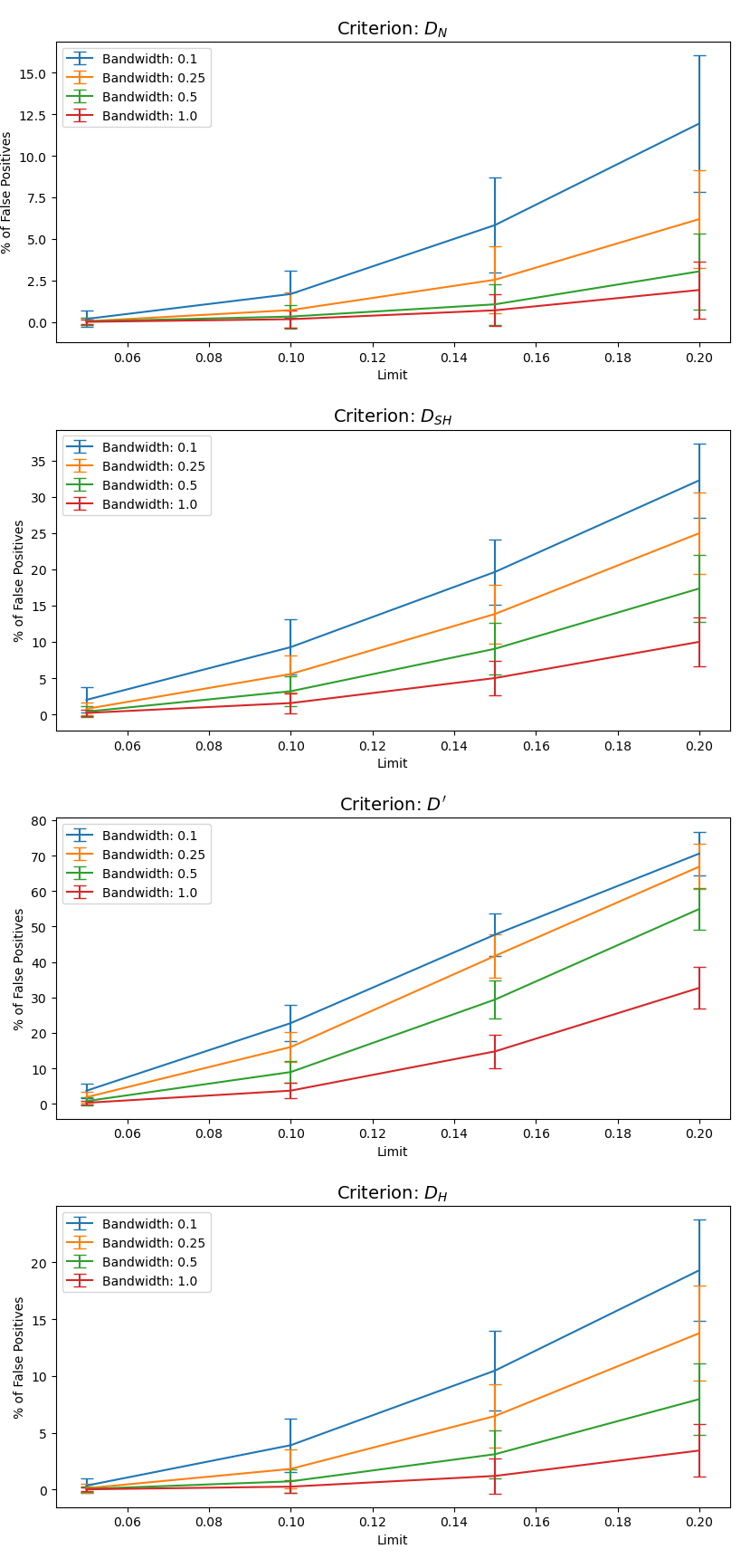}
    \caption{Percentage of false positives found for the $D_{SH}$, $D'$, $D_{H}$, $D_{N}$ for various bandwidth values and D-value limits.}
    \label{fig:false_positive_percent}
\end{figure}

The resulting false positive \% rate given different bandwidth parameters for the KDEs is seen in Fig.~\ref{fig:false_positive_percent}. For each similarity discriminant, the limiting value necessary to significantly decrease the false-positive rate varies widely, as the ranges of possible values are not identical. For example, $D_{N}$ achieves a false-positive rate of less than 5\% with a D-criterion value of $\sim$\,0.15 or less. Whereas, the $D'$ discriminant achieves similar levels of false positives when the limit is less than 0.05, i.e., if using the $D'$ one must set their criteria much lower. The $D_{H}$ discriminant attains a false-positive rate of $<$5\% around a limit of 0.1, and the classical $D_{SH}$ requires a limit of roughly 0.07 (depending on the bandwidth of the KDE used). However, these criteria and false-positive rates will vary depending on the fireball or meteor dataset being examined. 

\begin{figure}
	\includegraphics[width=\columnwidth]{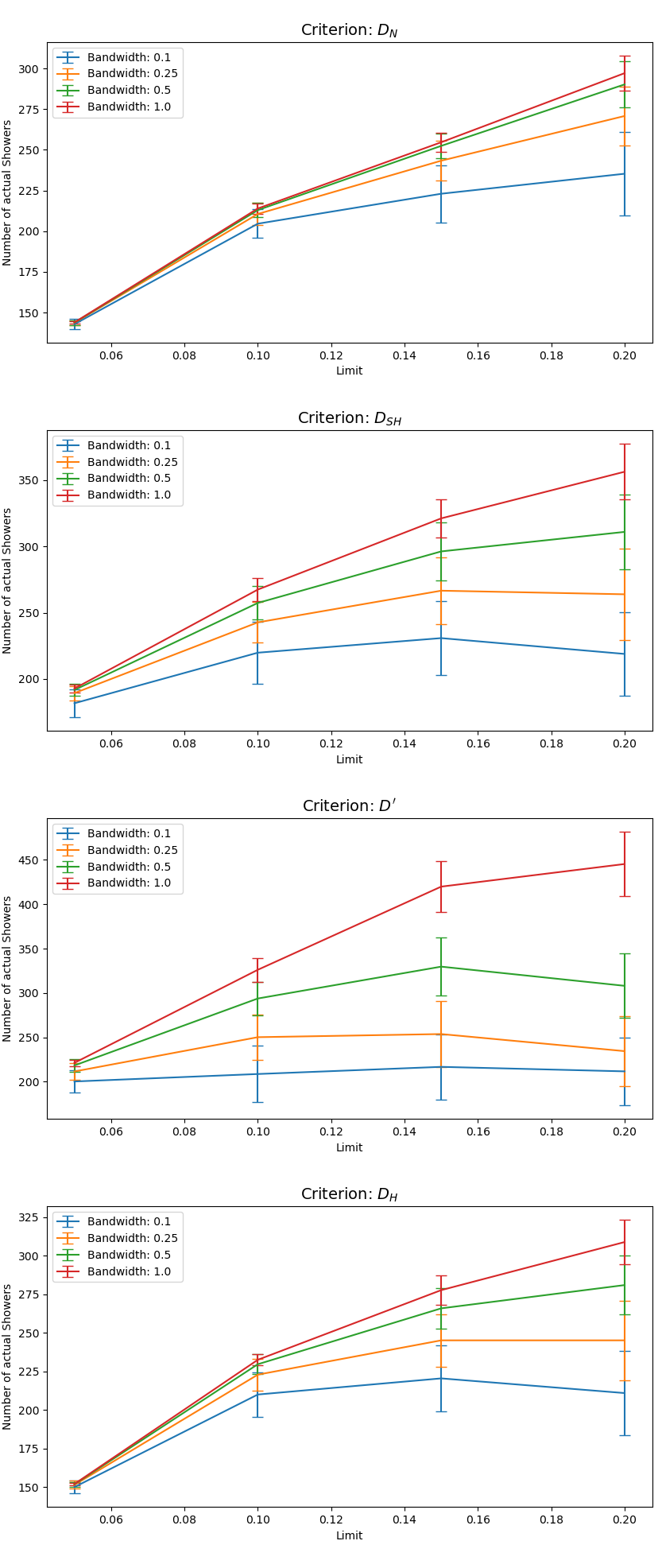}
    \caption{Estimate of the number of true showers associations for the EFN dataset, i.e., the number of fireballs that have D-values below the nominal limit for the observed dataset minus the estimated number of false positives for the given limit.}
    \label{fig:true_shower}
\end{figure}

If we calculate all the D-values for the EFN dataset and subtract the number of false positive estimates from the KDE analysis, we obtain Fig.~\ref{fig:true_shower}. This estimate of the number of ``true shower associations'' is meant to signify the number of fireballs that meet the D-criterion and are not a spurious association. The y-axis in this plot thus represents the estimate of true showers matches according to our KDE-produced sporadic meteor PDF. Despite the level of false positive differences between the similarity discriminants, interestingly, as the limit decreases, all four D values converge towards $\sim$150-200 shower-associated fireballs. This gives us confidence that the values we estimate are generalizable to any scalar similarity discriminant method. This also indicates that the level of established shower-associated fireballs within the EFN dataset is somewhere around 150-200 fireballs. Considering the size of their 2017-2018 dataset, this translates to approximately 18-25\% of the dataset. This of course does not account for minor showers on the large working list of the IAU MDC, but many of these may turn out to be not real, thus we did not consider them. 

In the studies by \citet{moorhead2016performance} and \citet{sugar2017meteor}, they used DBSCAN for explicit clustering of data into distinct groups of meteor showers, while the KDE is employed for estimating the underlying probability density function, which can be used to infer about the sporadic background and the likelihood of false positives in meteor shower identification. DBSCAN provides a more categorical interpretation (cluster vs. outlier), whereas KDE offers a probabilistic view of the data's distribution. A KDE may provide a more nuanced understanding of the data distribution, especially in cases where the boundaries between clusters are not clear-cut, but it requires careful selection of the bandwidth parameter. DBSCAN, on the other hand, is more straightforward in identifying dense regions but is sensitive to its core parameters. In summary, while DBSCAN is more focused on clustering and classifying individual data points into groups, KDE is used to estimate the overall distribution characteristics of the data, which is particularly useful in assessing the likelihood of false positives in meteor shower identification. Both of these methods have their unique advantages and can be complementary depending on the specific objectives of the analysis.

Lastly, it is important to note that this shower estimate is statistical in nature. Having a low D-value is important, as demonstrated by the decreasing false positive level as the limit decreases in Fig.~\ref{fig:false_positive_percent}. However, the meteors that meet the limiting requirement, irregardless of how low, could be spurious. For example, we estimate that between 6-11\% of the fireballs with a $D_{N}<0.2$ are false positives, but the false positives do not necessarily have to have the largest values in the subset. This is important to consider, especially when only applying these discriminants to a couple or even one fireball. 

\section{Conclusions}
Kernel Density Estimation (KDE), supplemented by rigorous data normalization techniques, provides a robust and flexible framework for estimating the sporadic background of meteor observations when the sporadic background is more significant than the meteor shower component. This approach allows for a more accurate assessment of meteor shower false positives, addressing the limitations inherent in traditional methods that rely heavily on the D-criterion. Our findings demonstrate that the optimal bandwidth, determined via cross-validation and grid search methods, plays a crucial role in achieving a balance between over- and under-smoothing of the data and obtaining an accurate false positive estimate.

Furthermore, our research underscores the importance of considering the statistical properties of meteoroid data, particularly the implications of sample size and observational uncertainties. By adopting a more nuanced and statistically sound methodology, we can enhance our understanding of meteor shower associations and their parent bodies. The generalisability of our approach, combined with its adaptability to different datasets and parameter sets can be used to create better statistics of near-Earth meteoroid population.

\section*{Acknowledgements}
This project has received funding from the European Union’s Horizon 2020 research and innovation programme under the Marie Skłodowska-Curie grant agreement No945298 ParisRegionFP.

\section*{Data Availability}

The EFN fireball data used in this study is available in the publication by \citet{borovivcka2022data}. 

The code developed to produce the results within study is also available at \url{https://zenodo.org/records/10406556}. 



\bibliographystyle{aa}
\bibliography{main} 








\label{lastpage}
\end{document}